\documentclass{elsart}

\usepackage{graphicx}

\usepackage{amssymb}

\begin{document}

\begin{frontmatter}

\title{The vortex depinning transition in untwinned YBaCuO using complex
impedance measurements.}
\author[crismat]{A. Pautrat, C. Goupil, Ch. Simon,}
\author[poznan]{B. Andrewjewski,}
\author[istec]{A.I. Rykov, S. Tajima.}
\address[crismat]{CRISMAT, ISMRA, UMR 6508 du CNRS, Caen, France.}
\address[poznan]{Institute of molecular physics, Poznan, Poland.}
\address[istec]{ISTEC, Tokyo, Japan.}
\begin{abstract}
We present surface impedance measurement of the vortex linear response in a large
untwinned YBCO crystal.  The depinning spectra obtained over a broad frequency range (100
Hz- 30 MHz) are those of a surface pinned vortex lattice with a free flux flow
resistivity (two modes response). The critical current in the "Campbell" like regime and
the flux flow resistivity in the dissipative regime are extracted. Those two parameters
are affected by the first order transition, showing that this transition may be related
to the electronic state of vortices.
\end{abstract}

\end{frontmatter}

The linear ac response of a pinned vortex lattice has been observed in the 60's in
conventional type II superconducting materials \cite{alais}. At low frequencies, say f
$\lesssim$ 1 KHz, a small ac field has an apparent penetration characterized by a static
regime without any loss (the Campbell regime). It has been recognized as a direct
consequence of the vortex pinning, and disagrees with first interpretations based on
thermally activated depinning \cite{campbell}. Note that this linear regime is not the
ohmic regime of a vortex lattice free from any pinning (the so called liquid phase). At
the same time, the high frequency response (f $\gtrsim$ 1-10 MHz) is that of a medium
with a free resistivity (skin effect). Those two regimes are separated by the depinning
frequency, and the shape of the whole spectrum allows to discriminate between different
types of pinned vortex states. One of the main interest of this method is that vortex
states can be studied even in the presence of a large critical current.
\begin{figure}
\begin{center}
\resizebox{!}{0.4\textwidth}{\includegraphics{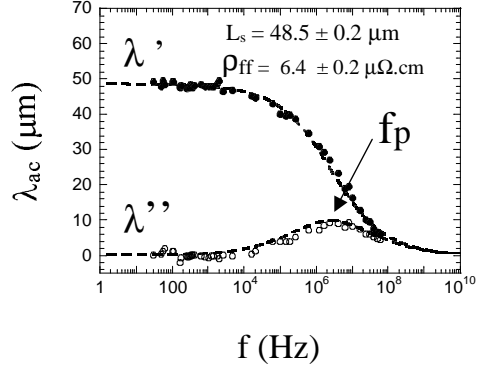}} \caption{Two modes response of
the pinned vortex lattice in untwinned YBaCuO ($T=88.6 K < T_{m}$, B//a-axis=6 T).}
\end{center}
\end{figure}
The principle of the experiment is to measure the ac penetration length, due to the
vortex response, in a small coil directly glued around the sample. The detailed
experimental procedure can be found in \cite{nous}. A typical depinning spectrum is shown
in the figure 1. It follows closely, over the whole frequency range, the two modes
 model of surface vortex pinning developed in \cite{sonin,nono}. At the same time, it
 was impossible to perform acceptable fits using bulk pinning models.  We conclude that the pinning in this YBaCuO sample
 is due to the surface roughness. We have also performed a precise study of the low frequency behavior close to the temperature of the first order transition. Within experimental accuracy, we do not find any evidence of a thermally assisted depinning regime. As the depinning frequency is directly measured by this
 method, we can identify the pure skin effect regime and measure the flux flow resistivity. A step in the flux flow resistivity is clearly identified at the temperature of the
 first order transition. At the same temperature, the low frequency penetration length
 diverges up to reach the full penetration (no pinning anymore). The first order transition affects both
 the critical current and the flux flow resistivity. It could suggest that this
 transition is linked to the pairing symmetry \cite{volovik}.

\end{document}